\documentclass[11pt]{article}

\usepackage[final]{acl}

\usepackage{times}
\usepackage{latexsym}
\usepackage[T1]{fontenc}
\usepackage[utf8]{inputenc}
\usepackage{microtype}
\usepackage{inconsolata}
\usepackage{graphicx}
\usepackage{booktabs}
\usepackage{enumitem}
\usepackage{amsmath}
\usepackage{amsfonts}
\usepackage{multirow}

\usepackage{xcolor}
\usepackage[most]{tcolorbox}
\tcbuselibrary{listings,breakable}
\title{Emotion Understanding in Streaming Video with \\ Trajectory-Aware Reliability}

\author{
  \textbf{Qingsong Wang$^{1,2}$, Qigong Lei$^{1}$, Zitong Wang$^{2}$, Bohan Yu$^{2}$, Zhiang Dong$^{1}$}\\
  \textbf{Jian Liu$^{2,\dagger}$, Weiqiang Wang$^{2}$, Chang Yao$^{1,3,\dagger}$, Jingyuan Chen$^{1,\dagger}$}\\
  \textit{$^1$Zhejiang University}\\
  \textit{$^2$Ant Group}\\
  \textit{$^3$Innovation and Management Center,}\\
  \textit{School of Software Technology (Ningbo), Zhejiang University}\\
\texttt{\{wqsong, changy, jingyuanchen\}@zju.edu.cn, rex.lj@antgroup.com}
\\
  $^{\dagger}$Corresponding authors.
}
\begin{document}
\maketitle
\begin{abstract}
Video emotion understanding is commonly studied as an offline classification problem, where the complete video segment is available before prediction.
Real-time interaction, however, requires emotion decisions from incomplete and evolving evidence.
This paper studies streaming video emotion understanding as a reliability-aware decision process over evolving emotion beliefs.
In this setting, a single confident prefix prediction can still be unreliable when the underlying belief trajectory is unstable or repeatedly switches across emotion classes.
We propose TRACE, a trajectory-aware reliability framework that forms low-latency emotion beliefs from streaming audio prefixes, estimates reliability from confidence, entropy, stability, and class-switching patterns, and selectively invokes contextual belief reinterpretation with visual, textual, and neighboring-utterance evidence.
TRACE keeps stable cases in the low-latency online pathway while allocating stronger multimodal reasoning to uncertain cases that remain ambiguous.
Experiments on StreamMER, MELD, and MER2024 show that TRACE improves the accuracy-cost trade-off, retaining most full-context gains while reducing unnecessary contextual reasoning.
Our code is available at \url{https://github.com/APTX574/Trace}.
\end{abstract}

\section{Introduction}

Video emotion understanding is fundamental to affective computing, dialog systems, and human-computer interaction~\cite{zenkri2026probingembodiedllmshigher,wang_iclr,pan2023review,sun2021multimodal}. Existing multimodal emotion recognition methods typically model facial expressions, acoustic cues, and linguistic content jointly, achieving substantial progress~\cite{cheng2024emotionllama,huang2025emotionqwen,yang2025omni,wang_emnlp}. However, most of these methods assume an offline setting where the complete video segment is available. In real-time interactive scenarios, such as embodied intelligence, online mental health support, and remote meeting analysis~\cite{zenkri2026probingembodiedllmshigher,astrin2026carecounseloralignedresponseengine,zhang2026generativeaitwotieredonline,chu-etal-2024-reconstruct}, a system must perceive user emotions dynamically as the interaction unfolds, rather than delaying its analysis until the full video segment concludes.

\begin{figure}[t]
\centering
\includegraphics[width=\columnwidth]{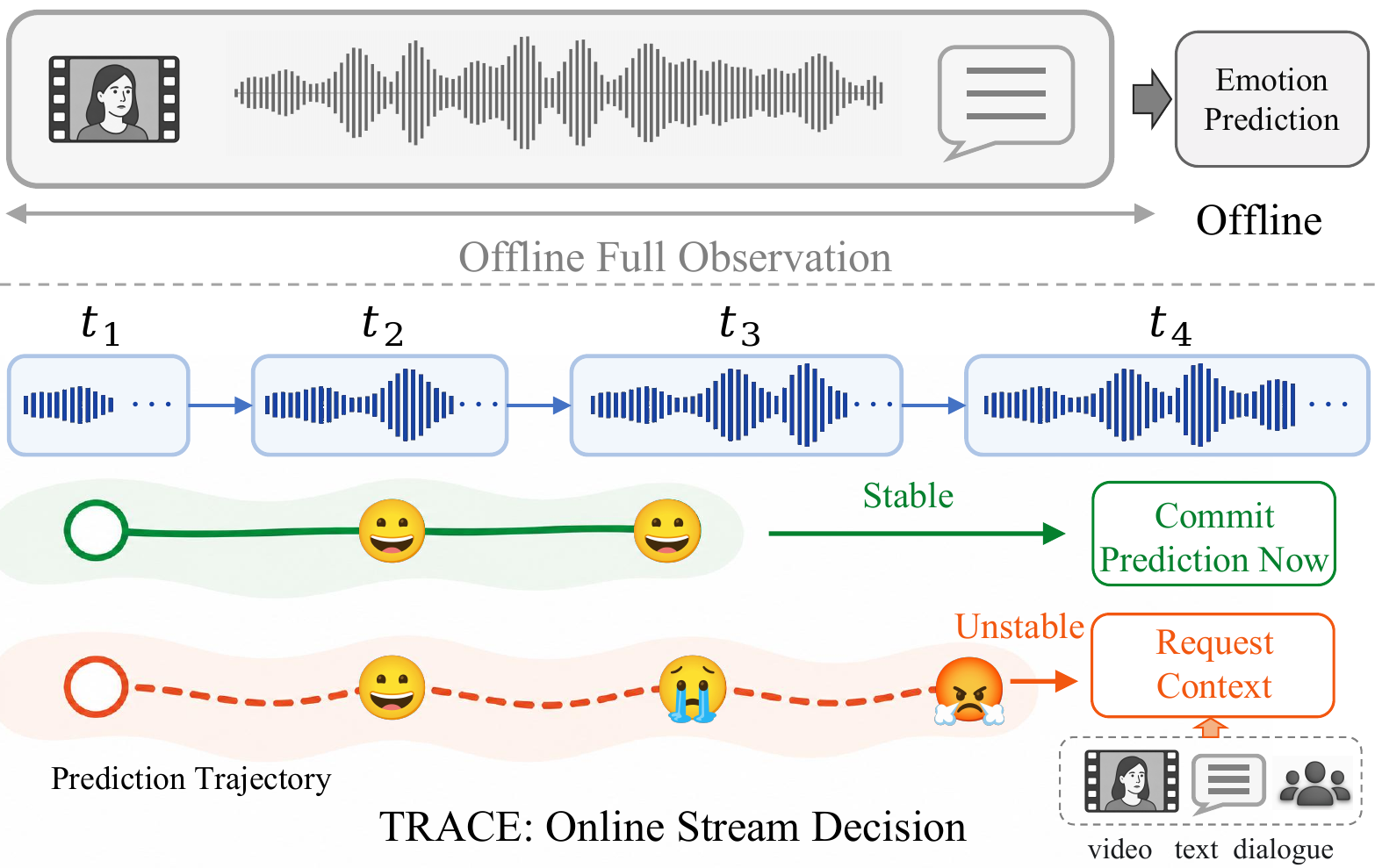}
\caption{\textbf{TRACE for streaming emotion understanding.}
Unlike offline emotion recognition, which predicts after full observation, TRACE tracks the evolving emotion belief during streaming inference.
Stable trajectories support immediate online commitment, whereas unstable trajectories trigger contextual reinterpretation with video, text, and dialogue evidence.}
\vspace{-10pt}
\label{fig:intro}
\end{figure}

Although existing streaming video understanding methods~\cite{liu2026efficientstreamingvideounderstanding,he2026videoodysseybenchmarkultralongcontextomnimodal,luo2026stsimdiffbalancingspatiotemporalsimilarity} can recognize actions, events, or scenes in real time, they generally wait for an event to fully conclude before making a decision~\cite{kang2024actionswitch,eyzaguirre2024streaming,song2024online}. Furthermore, they treat predictions at different time steps as independent static outputs. These methods translate poorly to streaming emotion understanding, which possesses three distinct characteristics. First, emotional expression is often anticipatory: even before an utterance or segment is fully observed, early audio, facial, or semantic cues may already contain sufficient evidence. Second, the prediction for the same utterance changes dynamically as new information arrives during streaming inference. The temporal trajectory of these changes reflects not only the current classification state, but also the process by which emotional evidence accumulates, beliefs stabilize, and uncertainty remains. In addition, different modalities exhibit asymmetric arrival patterns and decision value in streaming scenarios. Compared with vision and text, which often provide contextual supplementation and semantic correction, audio carries rich low-latency cues such as intonation, pauses, and rhythm. It is therefore more suitable for incremental processing and serves as a natural primary signal for low-latency belief tracking.

Based on these observations, this paper argues that \textbf{Streaming Video Emotion Understanding} should not be reduced to independent classification over multiple truncated prefixes. Instead, it should be formulated as a dynamic process in which an emotion belief is continuously formed, evaluated, and revised over time. This process requires the system to answer two core questions: whether the evidence accumulated so far is sufficient for low-latency prediction, and whether richer context is needed for joint reasoning. For samples whose beliefs converge quickly, the model should output predictions immediately to avoid unnecessary waiting. For samples whose beliefs keep fluctuating, remain insufficiently confident, or contain ambiguity, the model should defer the decision and introduce additional evidence for revision when needed.

Motivated by these observations, we propose TRACE, a trajectory-aware reliability framework for streaming video emotion understanding. TRACE contains three core components. As shown in Figure~\ref{fig:intro}, TRACE shifts emotion understanding from offline full-observation prediction to online stream decision-making. Online Prefix Belief Formation continuously estimates the emotion distribution and confidence state of the current segment from audio prefixes, producing an evolving belief trajectory over time. Trajectory-Calibrated Reliability Estimation adaptively determines whether the current prediction is reliable according to the confidence, stability, and trend of this trajectory, and decides whether the system should commit to the current belief, continue accumulating evidence, or request contextual reinterpretation. Contextual Belief Reinterpretation is activated only when the reliability estimation judges the current belief to be insufficient. It performs deeper cross-modal reinterpretation of the current segment by fusing visual cues, textual semantics, neighboring-segment context, and the prefix belief trajectory.

\begin{figure*}[t]
\centering
\includegraphics[width=1\linewidth]{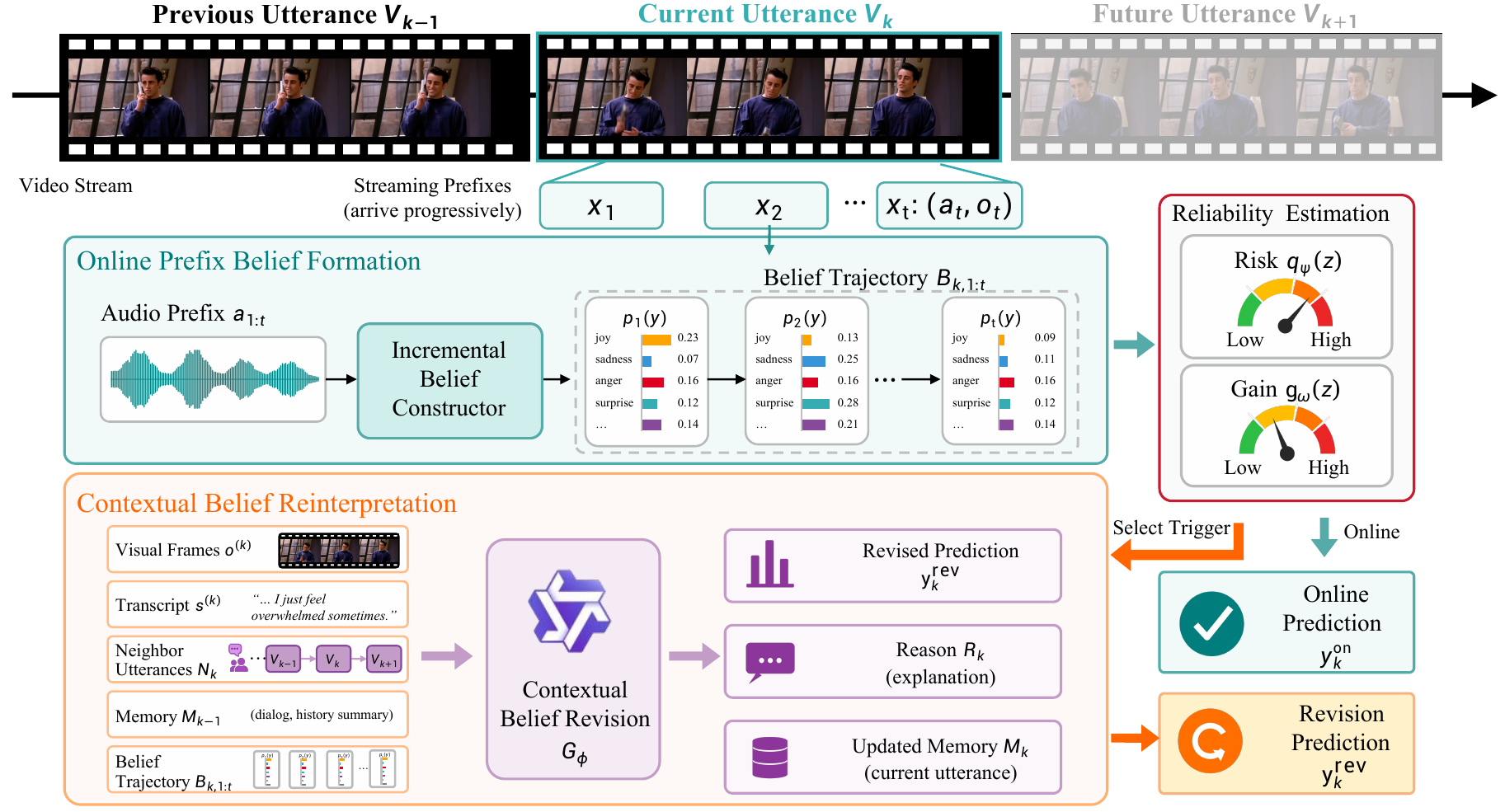}
\caption{\textbf{Overview of TRACE.} The framework first forms prefix-conditioned emotion beliefs from streaming audio, estimates trajectory-calibrated reliability from the evolving belief trajectory, and selectively performs contextual belief reinterpretation for uncertain samples.}
\vspace{-12pt}
\label{fig:overview}
\end{figure*}

Our main contributions are as follows:
\begin{itemize}[itemsep=0pt,topsep=2pt,parsep=0pt,leftmargin=*]
\item We define the task of streaming video emotion understanding and introduce StreamMER, a dataset designed to evaluate emotion recognition under partial and progressively revealed observations.
\item We propose a framework based on asymmetric evidence trajectories, which uses low-latency audio prefixes to model dynamic emotion belief evolution and adaptively decides when to invoke costly cross-modal reasoning.
\item Experiments on emotion understanding datasets show that our method reduces response latency and cross-modal reasoning calls while preserving prediction performance. 
\end{itemize}

\section{Related Work}

\textbf{Streaming video understanding.}
Recent video-language models~\cite{liang2026storminternalizedmodelingspatialtemporal,luo2026stsimdiffbalancingspatiotemporalsimilarity} have begun to address real-time and long-form video understanding. Methods such as BOLT~\cite{liu2025bolt} and VLog~\cite{lin2025vlog} improve long-video understanding through frame selection or retrieval-based video representation, while Flash-VStream~\cite{zhang2025flash} and StreamingVLM~\cite{xu2510streamingvlm} further target efficient understanding over continuous video streams. Dispider~\cite{qian2025dispider} and ViSpeak~\cite{fu2025vispeak} study interactive streaming scenarios, where the model needs to perceive ongoing visual inputs and respond in real time. However, these works mainly focus on general video events, actions, or user instructions, and usually do not model how an emotion prediction changes under partial observation. In contrast, our work studies streaming video emotion understanding, where the system tracks evolving emotion beliefs and decides whether to commit early, wait for more evidence, or request contextual reinterpretation.

\textbf{Video emotion understanding.}
Recent multimodal emotion models extend large language models to emotion recognition and reasoning. Emotion-LLaMA~\cite{cheng2024emotionllama}, AffectGPT~\cite{lian2025affectgpt}, EmoLLM~\cite{yang2024emollm}, Emotion-Qwen~\cite{huang2025emotionqwen}, and Omni-Emotion~\cite{yang2025omni} show that instruction tuning, emotion-specific datasets, facial cues, acoustic cues, and multimodal reasoning can improve video emotion understanding. These methods provide strong full-context emotion recognition ability, but they generally assume that the complete video, audio, text, or dialogue context is available before prediction. This offline setting is different from real-time interaction, where emotion judgments must often be made before an utterance is fully observed. TRACE is complementary to these models: it first forms low-latency emotion beliefs from streaming audio prefixes, and only invokes richer visual, textual, and contextual reasoning when the belief trajectory is unreliable. Visual representation quality is also crucial for multimodal reasoning. Frequency-aware visual encoders have shown improved understanding under challenging visual distributions \citep{wang_cvpr}.

\section{Method}

\subsection{Problem Setup and Framework Overview}
\label{sec:setup}

Given a video $V$, we apply voice activity detection to segment it into an ordered sequence of utterance-level units for streaming emotion analysis:
\begin{equation}
\mathcal{V} = \{V_1, V_2, \ldots, V_N\},
\end{equation}
where $V_k$ denotes the $k$-th utterance segment and its corresponding emotion analysis unit. 

In the streaming setting, models needs to predict the emotion of the current utterance before observing it completely. For utterance $V_k$, we denote the multimodal prefix observed up to time step $t$ as:
\begin{equation}
V_{k,t} = \{v_{k,1}, \ldots, v_{k,t}\}, \quad v_{k,t} = \{a_t, o_t\},
\end{equation}
where $a_{k,t}$ and $o_{k,t}$ denote the audio and visual information arriving at time step $t$, respectively.

When predicting the emotion of $V_k$, the system can access the historical utterances
$\mathcal{V}_{1:k-1} = \{V_1, \ldots, V_{k-1}\}$ and the current prefix $V_{k,t}$. However, it cannot access future content within $V_k$ or any subsequent utterances $V_{k+1}, \ldots, V_N$. Therefore, unlike conventional offline emotion recognition, streaming emotion understanding assumes that the complete utterance, transcript, and multimodal context are unavailable at prediction time.

Formally, the model continuously estimates the emotion distribution of the current utterance as:
\begin{equation}
p\left(y_k \mid \mathcal{V}_{1:k-1}, V_{k,t}\right),
\end{equation}
and dynamically updates its prediction as new evidence arrives. 

As shown in Figure~\ref{fig:overview}, TRACE separates low-latency Online Prefix Belief Formation from selective Contextual Belief Reinterpretation. It tracks emotion beliefs from audio prefixes and invokes visual, textual, and neighboring-utterance context only when the belief trajectory remains unreliable, thereby reducing overall inference latency.

\subsection{Online Prefix Belief Formation}
\label{sec:belief}
We use Qwen2.5-Omni-3B to conduct Online Prefix Belief Formation over streaming inputs. Audio prefixes serve as the primary evidence source because audio signals arrive continuously and provide low-latency emotional cues, such as intonation, pauses, speaking rate, and energy changes.

At each time step $t$, the model takes as input a sequence composed of the query and the current audio prefix, denoted as $[q, a_{1:t}]$, and outputs logits over the emotion label space. We then apply the softmax function to compute the probability of each emotion label $y \in \mathcal{Y}$, which is defined as the emotion belief distribution at the current time step:
\begin{equation}
p_{k,t}(y)=\frac{\exp(\ell_t(y))}{\sum_{y'\in\mathcal{Y}}\exp(\ell_t(y'))},
\end{equation}
where $\ell_t(y)$ denotes the output logit for label $y$. The online model's predicted label at the current time step is defined as:
$\hat{y}^{\mathrm{on}}_{k,t}$.

It is worth noting that although the model predicts a label $\hat{y}_{k,t}$ at each time step, these predictions are used only to construct the belief trajectory $\mathcal{B}_{1:t}=\{p_1,p_2,\ldots,p_t\}$ and are not concatenated into the input sequence of the next time step. Thus, subsequent belief updates are always driven by newly arrived audio evidence, rather than by an autoregressive accumulation of the model's earlier predictions. This avoids repeatedly encoding the full audio history and allows the model to exploit the KV cache, meeting the efficiency requirements of online inference.

During training, we apply supervision to all audio prefixes and use the emotion label $y_k$ of the current utterance as the training target. Since very early prefixes may contain weak emotional evidence, we weight different time steps according to the observation ratio. Let $T_k$ denote the number of prefix steps for the current utterance, and let $r_t=t/T_k$ denote the observation ratio of the $t$-th prefix. The training objective of the streaming belief estimator is defined as
\begin{equation}
\mathcal{L}_{\mathrm{IBC}}(V_k,y_k)=
\frac{
\sum_{t=1}^{T_k}
r_t\,\mathrm{CE}(p_{k,t},y_k)
}{
\sum_{t=1}^{T_k}r_t
},
\end{equation}
where $\mathrm{CE}(\cdot,\cdot)$ denotes the cross-entropy loss. This objective encourages the model to form usable emotion beliefs from early prefixes, while assigning higher supervision weights to prefixes closer to the complete segment.

\subsection{Trajectory-Calibrated Reliability Estimation}
\label{sec:policy}
The core of online decision making is to determine whether the current belief is already reliable. The confidence of a single prefix only describes the sharpness of the current distribution, but it does not indicate whether the distribution is gradually formed from stable evidence or induced by short-term fluctuations.

Hence, we extract a trajectory state $z_{k,t}=\phi(\mathcal{B}_{k,1:t})$ from the belief trajectory, and use it to estimate both the risk of directly committing the current prediction and the potential gain of contextual reinterpretation. The state $z_{k,t}$ includes the current distribution, confidence changes, entropy changes, class switching patterns, distributional stability, and the observation ratio $t/T_k$. More details and analyses are reported in the Appendix~\ref{sec:feature-details}.

Based on $z_{k,t}$, we train a two-head reliability model. The risk head $q_\psi(z_{k,t})$ estimates the probability that the current belief prediction is incorrect, while the gain head $g_\omega(z_{k,t})$ estimates whether Contextual Belief Reinterpretation, whose output $\hat{y}^{\mathrm{rev}}_k$ is described in Section~\ref{sec:revision}, can correct the current error. The risk label is defined as:
\begin{equation}
e_{k,t}=\mathbb{I}(\hat{y}^{\mathrm{on}}_{k,t}\neq y_k),
\end{equation}
with $\mathbb{I}(\cdot)$ denoting the indicator function, and the gain label is defined as:
\begin{equation}
b_{k,t}=\mathbb{I}(\hat{y}^{\mathrm{on}}_{k,t}\neq y_k \land \hat{y}^{\mathrm{rev}}_k=y_k).
\end{equation}
The heads estimate whether the belief is wrong and whether additional context can correct it.

\begin{table*}[t]
\centering
\small
\caption{Overall emotion recognition results on three datasets. Values are percentages. F1 denotes Weighted-F1. TRACE achieves performance close to full contextual reinterpretation while invoking reinterpretation for only 55.58\%, 54.63\%, and 58.21\% of samples on StreamMER, MELD, and MER2024, respectively.}
\label{tab:main-results}
\resizebox{\textwidth}{!}{
\begin{tabular}{lccccccccc}
\toprule
\multirow{2}{*}{Method} &
\multicolumn{3}{c}{StreamMER} &
\multicolumn{3}{c}{MELD} &
\multicolumn{3}{c}{MER2024} \\
\cmidrule(lr){2-4}\cmidrule(lr){5-7}\cmidrule(lr){8-10}
& Acc. & F1 & UAR & Acc. & F1 & UAR & Acc. & F1 & UAR \\
\midrule
Gemini-3-Flash & 57.98 & 56.68 & 58.59 & 50.46 & 50.34 & 36.75 & 77.07 & 78.17 & 75.71 \\
Gemini-3-Pro & 63.08 & 62.84 & 66.91 & 61.34 & 62.27 & 45.62 & 86.82 & 86.95 & 84.68 \\
Emotion-Llama & 31.26 & 29.18 & 27.18 & 28.85 & 24.10 & 33.61 & 43.28 & 45.73 & 50.99 \\
AffectGPT & 53.66 & 49.86 & 48.51 & 55.65 & 51.33 & \textbf{56.67} & 78.80 & 82.46 & 75.74 \\
Emotion-Qwen & 59.51 & 58.39 & 48.45 & 51.76 & 52.04 & 38.51 & 82.12 & 82.50 & 80.77 \\
ViDEmo & 55.46 & 58.10 & 32.25 & 57.13 & 59.20 & 46.05 & 66.55 & 66.92 & 66.98 \\
\midrule
Online belief formation only & 59.87 & 54.56 & 52.58 & 58.94 & 56.08 & 38.67 & 81.18 & 80.09 & 75.29 \\
Full contextual reinterpretation & \textbf{70.18} & \textbf{70.01} & \textbf{66.47} & \textbf{70.30} & 68.90& 51.51 & \textbf{87.42} & \textbf{87.09} & \textbf{85.71} \\
TRACE & 69.47 & 68.10 & 63.80 & 68.28 & \textbf{70.14} & 41.42 & 85.46 & 85.11 & 83.04 \\
\bottomrule
\end{tabular}}
\vspace{-10pt}
\end{table*}

The two-head model shares an MLP encoder and uses two sigmoid heads to output $q_\psi$ and $g_\omega$, respectively. The training objective is the sum of two binary cross-entropy losses:
\begin{equation}
\begin{aligned}
\mathcal{L}_{\mathrm{rel}}
=
\sum_{k,t} w_{k,t}
\Big[
&
\mathrm{BCE}(q_\psi(z_{k,t}),e_{k,t})
\\
&+
\mathrm{BCE}(g_\omega(z_{k,t}),b_{k,t})
\Big],
\end{aligned}
\end{equation}
where $w_{k,t}$ emphasizes samples that truly benefit from reinterpretation, reduces the influence of samples where reinterpretation instead causes errors.

The reliability estimates are then used as part of a learned online decision policy. We feed the trajectory state $z_{k,t}$, the risk estimate $q_\psi(z_{k,t})$, and the gain estimate $g_\omega(z_{k,t})$ into a lightweight policy head $f_\theta$, which predicts the action distribution at the current time step:
\begin{equation}
\mathbf{s}_{k,t}
=
\mathrm{softmax}
\left(
f_\theta
\left(
z_{k,t},
q_\psi(z_{k,t}),
g_\omega(z_{k,t})
\right)
\right),
\end{equation}
where $\mathbf{s}_{k,t}$ is the action distribution over committing the current prediction, waiting for more evidence, and requesting contextual reinterpretation. During online inference, the selected action is:
\begin{equation}
a_{k,t}
=
\arg\max_{a\in\{\mathrm{commit},\mathrm{wait},\mathrm{reinterpret}\}}
\mathbf{s}_{k,t}(a).
\end{equation}

\subsection{Contextual Belief Reinterpretation}
\label{sec:revision}

When Trajectory-Calibrated Reliability Estimation judges the current audio belief as unreliable, the system enters Contextual Belief Reinterpretation.
For such cases, audio prefixes may remain unstable or ambiguous, and the correct emotion may depend on facial expressions, spoken content, or preceding context.
Thus, we treat cross-modal reasoning as a selectively requested reinterpretation process rather than a default offline classification step.

If the policy $f_\theta$ selects reinterpretation, the system invokes the cross-modal model $G_\phi$.
Unlike Online Prefix Belief Formation, $G_\phi$ uses richer evidence at the cost of additional latency to re-evaluate high-risk utterances.
For a selected utterance $V_k$, it receives the multimodal information of the current utterance, the preceding context, historical memory, and the formed belief trajectory:
\begin{equation}
\mathcal{C}_k=
\{a^{(k)},o^{(k)},s^{(k)},V_{k-1},M_{k-1},\mathcal{B}_{k,1:t}\},
\end{equation}
where $a^{(k)}$ denotes the audio of the current utterance, $o^{(k)}$ denotes the visual observations sampled from the current video, $s^{(k)}$ denotes textual semantic information, $V_{k-1}$ denotes the available context from the preceding utterance, $M_{k-1}$ denotes the previous contextual memory, and $\mathcal{B}_{k,1:t}$ denotes the prefix belief trajectory formed before reinterpretation is requested. In implementation, visual observations are sampled from the original video at 2 fps and processed by the visual encoder of Qwen2.5-Omni-7B~\footnote{https://arxiv.org/abs/2503.20215}.
We explicitly feed $\mathcal{B}_{k,1:t}$ into the reinterpretation model because the belief trajectory reveals why Online Prefix Belief Formation produces an unreliable judgment, such as frequent class switching or long-term competition among multiple classes. Therefore, the reinterpretation model does not simply reclassify the current utterance, but performs contextual evidence supplementation and re-judgment based on the already formed dynamic belief.
The output of the reinterpretation model is:
\begin{equation}
(\hat{y}^{\mathrm{rev}}_k,R_k,M_k)=
G_\phi(\mathcal{C}_k),
\end{equation}
where $\hat{y}^{\mathrm{rev}}_k$ is the reinterpreted emotion classification result, $R_k$ is the classification rationale, and $M_k$ is the structured summary of the current utterance together with the updated contextual memory. We further ablate the roles of rationale generation and contextual memory summary in Appendix~\ref{sec:rationale-summary-ablation}.

Therefore, the final prediction is determined by the action selected by the online policy. If the policy chooses immediate commitment, the system uses the current belief prediction $\hat{y}^{\mathrm{on}}_{k,t}$. If the policy chooses waiting, the system continues to accumulate new prefixes. If the policy chooses contextual reinterpretation, the system uses the reinterpreted result $\hat{y}^{\mathrm{rev}}_k$. Through this design, the system keeps samples supported by stable audio trajectories in Online Prefix Belief Formation, and introduces stronger visual, textual, and contextual reasoning only for samples whose trajectories require additional evidence.

\begin{figure}[t]
\centering
\includegraphics[width=\columnwidth]{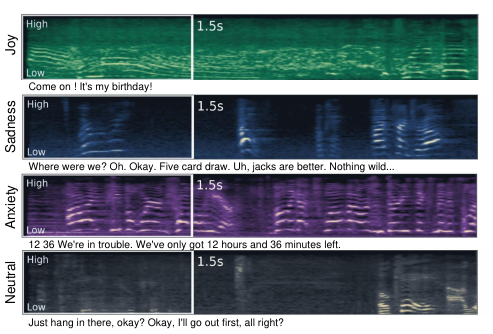}
\caption{Acoustic evidence in early audio prefixes. Spectrogram examples show that different emotion categories exhibit distinguishable acoustic patterns within short prefix observations.}
\vspace{-15pt}
\label{fig:prefix_show}
\end{figure}

\section{Experiments}
\label{sec:experiments}

\subsection{Experimental Setup}

\paragraph{Dataset.}
We construct StreamMER, a streaming-oriented video emotion dataset from the first two seasons of \textit{Friends}. Unlike conventional utterance-level emotion datasets that treat each utterance as an isolated sample, StreamMER preserves local dialogue context and scene-level interaction. We first segment the episodes into coherent narrative units of approximately 1.5 minutes using the scripts, subtitle files, and video content. Each segment contains a short conversational scene with surrounding dialogue, speaker interactions, and local emotional context.
For each utterance in a segment, we generate an initial emotion annotation with the help of Gemini-3.1-Pro~\footnote{\href{https://deepmind.google/models/model-cards/gemini-3-1-pro}{Gemini-3.1-Pro model card}}. The resulting StreamMER dataset contains 8,111 training utterances and 815 test utterances. We provide construction details and dataset statistics in Appendix~\ref{sec:dataset-appendix}, and annotation prompts in Appendix~\ref{app:prompt-templates}.

In our experiments, StreamMER is evaluated under the streaming protocol: Online Prefix Belief Formation only observes the arrived audio prefix, while Contextual Belief Reinterpretation is requested selectively and its additional latency and inference cost are counted. We also conduct experiments on MELD~\cite{poria-etal-2019-meld} and MER2024~\cite{lian2024mer} as offline multimodal emotion understanding benchmarks, since their standard settings and existing baselines do not provide strict streaming observations.

\begin{figure}[t]
\centering
\includegraphics[width=\columnwidth]{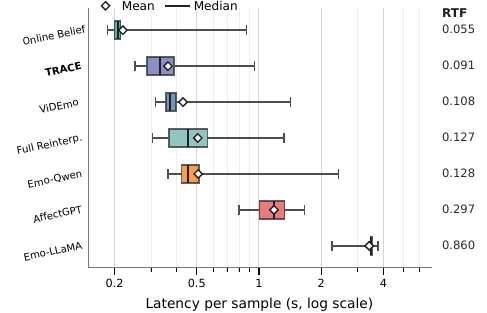}
\caption{Latency distribution and real-time factor (RTF) of different inference configurations.}
\vspace{-18pt}
\label{fig:latency}
\end{figure}

\noindent\textbf{Metrics.}
We report Accuracy, Weighted-F1, and UAR for emotion recognition. For streaming evaluation on StreamMER, we further report Average Decision Time, Normalized Decision Ratio, Reinterpretation Rate, and real-time factor (RTF). Reinterpretation Rate denotes the proportion of samples assigned to Contextual Belief Reinterpretation.

\noindent\textbf{Baselines.}
We compare TRACE with open-source multimodal emotion models, including Emotion-LLaVA~\cite{cheng2024emotionllama}, Affect~\cite{lian2025affectgpt}, Emotion-Qwen~\cite{huang2025emotionqwen}, and ViDEmo~\cite{zhang2026videmo}, as well as closed-source multimodal LLMs such as Gemini-3-Flash~\footnote{\href{https://deepmind.google/models/model-cards/gemini-3-flash/}{Gemini-3-Flash model card}} and Gemini-3-Pro. On StreamMER, we further compare streaming decision baselines, including Full-Prefix Belief Formation, Universal Contextual Reinterpretation, Rule-Based Reliability Estimation, Trajectory-Calibrated Reliability Estimation, and Oracle Reinterpretation Policy. 
Oracle Reinterpretation Policy uses ground-truth labels to select between belief-formation and reinterpretation predictions and is reported only as an upper bound.

\noindent\textbf{Implementation details.}
Online Prefix Belief Formation uses Qwen2.5-Omni-3B, and Contextual Belief Reinterpretation uses Qwen2.5-Omni-7B. For reinterpretation, video frames are sampled at 2 fps and resized to $448\times448$. Trajectory-Calibrated Reliability Estimation uses trajectory features such as confidence, entropy, class margin, label switching, JS divergence, top-1 persistence, and observation ratio. More details are provided in Appendix~\ref{sec:feature-details}, and inference prompts are provided in Appendix~\ref{app:prompt-templates}.

\subsection{Overall Results}
\label{sec:main-results}

Table~\ref{tab:main-results} reports the overall emotion recognition results on the three datasets.
Full contextual reinterpretation achieves the best overall performance on all three datasets, suggesting that complete multimodal information and contextual evidence provide stable value for emotion recognition.
Compared with general multimodal baselines, this model incorporates preceding conversational context on StreamMER and MELD, in addition to the complete audio, visual, and textual information of the current utterance.
It therefore better handles emotion samples that depend on context.
This indicates that contextual reinterpretation provides an effective complement for samples with insufficient information or unstable predictions in streaming emotion understanding.

\begin{figure}[t]
\centering
\includegraphics[width=\columnwidth]{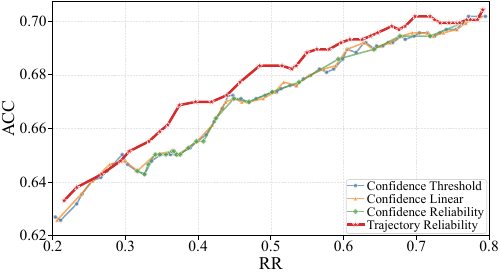}
\caption{\textbf{Accuracy across contextual reinterpretation rates.} Trajectory Reliability shows a better accuracy-cost trade-off than confidence-based policies.}
\vspace{-15pt}
\label{fig:acc-rr}
\end{figure}

At the same time, online belief formation only already outperforms or approaches general multimodal baselines in several settings.
For example, on StreamMER, it achieves 59.87\% Accuracy, outperforming Affect and ViDEmo and approaching Emotion-Qwen.
This result suggests that streaming audio prefixes contain useful emotion cues and can support effective early belief formation.
Figure~\ref{fig:prefix_show} further supports this observation: early audio prefixes exhibit emotion-relevant spectral and temporal patterns.
As these prefix predictions evolve over time, they naturally form belief trajectories that can be used to judge prediction stability. Additional trajectory analyses are provided in Appendix~\ref{sec:belief-trajectory-analysis}. A complete set of prefix spectrogram visualizations is provided in Appendix~\ref{sec:prefix-visualization}.

TRACE preserves most of the gains from full contextual reinterpretation while substantially reducing reinterpretation requests.
On StreamMER, TRACE improves Accuracy from 59.87\% to 69.47\% over online belief formation only, while invoking contextual reinterpretation for only 55.58\% of samples.
On MER2024 and MELD, TRACE similarly uses reinterpretation for only 58.21\% and 54.63\% of samples, respectively, while remaining close to full contextual reinterpretation in Accuracy and F1.
These results show that TRACE improves the accuracy-cost trade-off by using belief trajectories to decide when to commit early and when to request stronger contextual reasoning.
The following sections further analyze this trade-off from latency, routing, and ablation perspectives.

\begin{figure*}[t]
\centering
\includegraphics[width=1\linewidth]{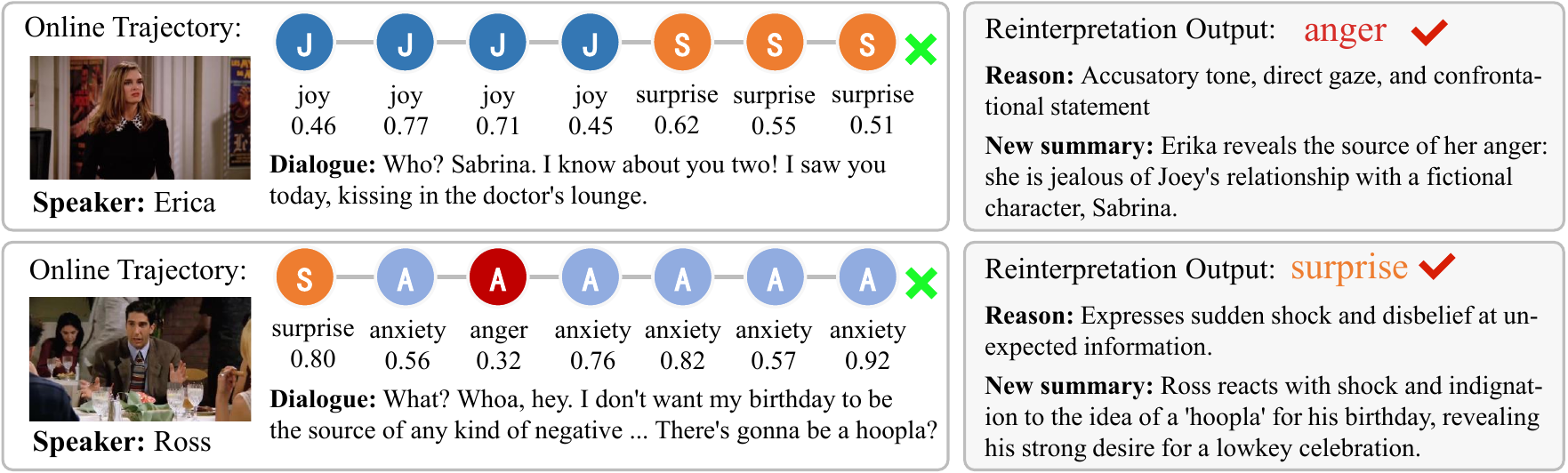}
\vspace{-5pt}
\caption{
\textbf{Case study of TRACE.}
The prefix-level belief trajectories reveal unstable online predictions, which trigger Contextual Belief Reinterpretation.
By incorporating visual, textual, and dialogue context, TRACE revises ambiguous audio-prefix beliefs into more reliable emotion predictions.
}
\vspace{-10pt}
\label{fig:case}
\end{figure*}

\begin{table}[t]
\centering
\small
\caption{\textbf{Streaming decision strategy comparison on StreamMER}. NDR denotes normalized decision ratio, RR denotes contextual reinterpretation rate.}
\label{tab:selective-reinterpretation}
\resizebox{\columnwidth}{!}{
\begin{tabular}{lccccc}
\toprule
Strategy & Acc.$\uparrow$ & W-F1$\uparrow$ & Avg. Time$\downarrow$ & NDR$\downarrow$ & RR$\downarrow$ \\
\midrule
Early Commit & 57.98 & 57.38 & \textbf{1.49} & \textbf{50.77} & - \\
Full Commit & 58.98 & 58.41 & 3.98 & 100.00 & - \\
Adaptive Commit & 59.65 & 58.99 & 3.10 & 87.66 & - \\
TRACE & 69.47 & 68.10 & 3.10 & 87.66 & 55.58 \\
Always Reinterp. & 70.18 & 70.03 & 4.05 & 100.00 & 100.00 \\
Oracle Reinterp. & \textbf{75.50} & \textbf{75.25} & - & 100.00 & \textbf{16.52} \\
\bottomrule
\end{tabular}}
\vspace{-10pt}
\end{table}

\subsection{Latency and Throughput}
\label{sec:latency}

Streaming emotion understanding requires timely prediction under real-time interaction constraints. We therefore compare the per-sample latency and real-time factor (RTF) of different model configurations under the same hardware setting.

As shown in Figure~\ref{fig:latency}, Online Prefix Belief Formation has the lowest cost, with an RTF of 0.055.
Full Contextual Belief Reinterpretation introduces higher latency, with an RTF of 0.127, showing that applying contextual multimodal reasoning to every sample is inefficient for streaming deployment. Larger multimodal emotion models further increase inference cost.
TRACE achieves an RTF of 0.091, remaining close to Online Prefix Belief Formation while being faster than full reinterpretation and larger-model baselines. This indicates that Trajectory-Calibrated Reliability Estimation limits Contextual Belief Reinterpretation to uncertain trajectories, preserving low latency for stable samples while allocating additional computation only when richer evidence is needed.

\subsection{Contextual Belief Reinterpretation}
\label{sec:selective-reinterpretation}

We further evaluate whether selective contextual reinterpretation provides a better trade-off between prediction accuracy, decision time, and contextual reasoning cost.
Here, a streaming decision strategy denotes how the system decides when to stop observing the current utterance and whether to use the online belief prediction or Contextual Belief Reinterpretation.
We compare TRACE with several strategies on StreamMER.
Early Commit commits at the first 1.5s prefix.
Full Commit waits until the full utterance but does not use reinterpretation.
Adaptive Commit follows the same waiting behavior as TRACE but disables reinterpretation.
Always Reinterp. applies reinterpretation to all samples.
Oracle Reinterp. uses ground-truth labels to select whether reinterpretation should be applied and is reported only as an upper bound.

Table~\ref{tab:selective-reinterpretation} shows that simply observing more audio is insufficient.
Early Commit reaches only 57.98 Accuracy, while waiting for the full utterance improves accuracy slightly to 58.98 with Full Commit.
Adaptive Commit reaches 59.65, indicating that many ambiguous cases cannot be resolved by audio-prefix accumulation alone.

Contextual reinterpretation brings clear gains but is costly when applied to all samples.
Always Reinterp. reaches 70.18 Accuracy with 100.00 RR.
In contrast, TRACE achieves 69.47 Accuracy and 68.10 W-F1 with only 55.58 RR, using the same average decision time as Adaptive Commit.
This shows that TRACE improves performance mainly by selectively invoking contextual reinterpretation, rather than by waiting longer.
The oracle result, 75.50 Accuracy with 16.52 RR, further suggests room for better reliability estimation.
\vspace{-5pt}

\begin{table}[t]
\centering
\small
\caption{Ablation of reliability features and estimation models. Acc. and MF1 denote Accuracy and Macro-F1. RR denotes the contextual reinterpretation rate.}
\label{tab:reliability-ablation}
\resizebox{\columnwidth}{!}{
\begin{tabular}{lccc}
\toprule
Reliability Method & Acc.$\uparrow$ & MF1$\uparrow$ & RR$\downarrow$ \\
\midrule
Confidence Linear & 66.87 & 64.30 & 69.33 \\
Confidence Reliability & 67.98 & \textbf{65.87} & 64.66 \\
Trajectory Linear & \textbf{68.83} & 65.79 & 60.12 \\
Trajectory Reliability & \textbf{68.83} & 65.40 & \textbf{55.71} \\
\bottomrule
\end{tabular}}
\vspace{-10pt}
\end{table}

\subsection{Ablation Studies}
\label{sec:ablations}

We conduct ablation studies to evaluate Trajectory-Calibrated Reliability Estimation by comparing single-prefix confidence with belief-trajectory features, and a linear selection rule with the proposed reliability model.

Table~\ref{tab:reliability-ablation} shows that trajectory features are more informative than confidence alone.
Under the linear rule, they improve Accuracy from 66.87 to 68.83 and reduce the reinterpretation rate from 69.33 to 60.12.
Under the reliability model, they improve Accuracy from 67.98 to 68.83 and reduce the reinterpretation rate from 64.66 to 55.71.
These results indicate that belief trajectories capture reliability patterns beyond single-prefix confidence.
Explicit reliability modeling further improves decision efficiency.
With confidence features, it improves Accuracy and Macro-F1 while reducing reinterpretation requests.
With trajectory features, it matches the Accuracy of the linear rule but requires fewer contextual reinterpretations.
Overall, belief-trajectory features and reliability estimation enable more cost-effective contextual reasoning under limited reinterpretation budgets.

We further examine the audio-first design through fast-path modality
ablations, acoustic-noise stress tests, cross-dataset evaluation, and a
detailed routing audit. These additional analyses are reported in
Appendix~\ref{app:robustness}.
\vspace{-5pt}

\subsection{Case Study}
\label{sec:case-study}

Figure~\ref{fig:case} shows two representative cases of TRACE.
In the first case, the audio-prefix belief shifts from \textit{joy} to \textit{surprise}, indicating an unstable trajectory.
With contextual reinterpretation, the model uses the confrontational dialogue, direct gaze, and accusatory tone to revise the prediction to \textit{anger}.
In the second case, the belief trajectory fluctuates among \textit{surprise}, \textit{anxiety}, and \textit{anger}.
After incorporating the spoken content and dialogue context, the model resolves the ambiguity and outputs \textit{surprise}.
These cases show that unstable belief trajectories can reveal when audio-prefix predictions are unreliable, and that selective contextual reinterpretation helps correct ambiguous online beliefs with richer multimodal evidence.

\section{Conclusion}

This paper presents TRACE, a trajectory-aware framework for streaming video emotion understanding. TRACE models emotion prediction as belief evolution under partial observation and uses belief trajectories to decide when to commit early or invoke contextual reinterpretation. Experiments on StreamMER and two public benchmarks show that TRACE improves the accuracy-cost trade-off, highlighting belief stability as an important signal for reliable real-time emotion understanding.

\section*{Acknowledgments}
This work was supported by the "Pioneer" and "Leading Goose" R\&D Program of Zhejiang (Grant No. 2025C02022), the Ant Group Research Fund, the Ningbo "Yongjiang Talent Program" Youth Innovation Project (Grant No. 2024A-156-G), and the Young Scientists Fund of the National Natural Science Foundation of China (Grant No. 62507040). In writing, we used ChatGPT for polishing.

\section*{Limitations}

TRACE improves the accuracy-cost trade-off for streaming video emotion understanding, but several limitations remain. First, its reliability estimator relies on trajectory patterns from calibration data and may be less reliable under substantial domain shifts. Second, contextual belief reinterpretation still introduces extra latency and computation when invoked, making TRACE less suitable for extremely strict real-time applications. Third, using audio prefixes as the primary low-latency signal may overlook early visual cues such as facial expressions or actions. Finally, StreamMER is built from scripted conversational videos, so future work should evaluate TRACE on more spontaneous and diverse real-world scenarios.

\section*{Ethics Statement}

StreamMER is constructed for academic research using publicly accessible video, subtitle, and script resources. We do not redistribute the original videos, subtitles, scripts, screenshots, or any copyrighted media content. The released dataset will contain only derived annotations and metadata, such as emotion labels, timestamps, speaker identifiers, context summaries, rationales, and non-reconstructive features when applicable. Users who wish to reproduce the benchmark must obtain the original media through lawful channels and comply with the corresponding content providers' terms. The annotations describe perceived emotions of fictional characters in scripted scenes and should not be interpreted as psychological assessment of real individuals.

\bibliography{custom}

\appendix

\section{Dataset Construction and Statistics}
\label{sec:dataset-appendix}

\paragraph{Data source and segmentation.}
StreamMER is built from the first two seasons of \textit{Friends}. We use the original videos together with script and subtitle files to segment each episode into coherent narrative units. Each unit preserves a locally complete interaction scene, including preceding dialogue state, speaker relations, and local emotional development. This design is important for streaming emotion understanding because the emotion of an utterance is often shaped by the immediately preceding conversational context.

\paragraph{Annotation protocol.}
For each segmented scene, we annotate utterances at the utterance level. The annotation process uses the video segment, aligned dialogue, and local script context. For each utterance, the annotator records the speaker, dialogue text, timestamp, emotion label, local context summary, rationale, audio evidence, and visual evidence. The rationale explains why the utterance is assigned to the target emotion, while the audio and visual evidence fields record cues such as prosody, pitch, speaking rate, facial expression, gaze, and body posture.

\paragraph{Human verification.}
Model-assisted data construction provides a scalable way to generate
structured supervision, while recent synthesis pipelines have also
highlighted the importance of explicit validation for maintaining data
fidelity \citep{chen2025toolforge}. Following this principle, we use
model-generated annotations only as drafts rather than final ground
truth.
The initial annotations are generated with the help of Gemini-3.1-Pro and then manually verified by three annotators. All annotators are graduate students in computer science with experience in multimodal machine learning or affective computing. Each annotator independently checks the initial label, timestamp, speaker, rationale, and audio/visual evidence against the video, audio, subtitle, and local script context. The final emotion label is determined by majority vote among the three annotators. When all three annotators disagree or when the evidence is judged insufficient, the case is discussed jointly and either assigned a consensus label or removed from the dataset if no reliable dominant emotion can be identified. We remove utterances that do not contain clear emotional content, are semantically incomplete, or are not useful for emotion understanding. We also correct wrong labels and check whether the emotion rationale is supported by the video, audio, and surrounding dialogue. Approximately 30\% of the initial Gemini-generated labels are modified during human verification, indicating that the model-generated labels are used only as annotation drafts rather than as final ground truth. Special attention is given to ambiguous cases such as sarcasm, teasing, mild anxiety, embarrassment, and neutral utterances, since these categories are easily confused without context.

\paragraph{Data release and copyright.}
StreamMER is constructed from copyrighted television episodes and associated subtitle/script resources. We do not redistribute the original videos, subtitles, scripts, screenshots, or any other copyrighted media content. The released research artifact contains only derived annotations and metadata needed to reproduce the benchmark, such as episode and timestamp references, utterance-level emotion labels, speaker identifiers, local context summaries, rationales, and precomputed non-reconstructive features when applicable. Users must obtain the original media through lawful channels and comply with the licenses and terms of the underlying content providers. The dataset is intended for academic research on streaming multimodal emotion understanding, and the annotations should not be interpreted as granting permission to redistribute or republish the underlying media.

\begin{table}[t]
\centering
\small
\caption{Statistics of StreamMER. Duration denotes the total duration of annotated utterance videos.}
\label{tab:dataset-statistics}
\begin{tabular}{lcccc}
\toprule
Split & Utterances & Clips & Episodes & Duration \\
\midrule
Train & 8,111 & 365 & 48 & 9.41 h \\
Test & 815 & 40 & 26 & 0.92 h \\
\bottomrule
\end{tabular}
\end{table}

\begin{table}[t]
\centering
\small
\caption{Duration statistics of StreamMER. Values are measured on annotated utterance videos.}
\label{tab:duration-statistics}
\resizebox{\columnwidth}{!}{
\begin{tabular}{lcccc}
\toprule
Split & Mean Utt. & Median Utt. & Mean Clip & Median Clip \\
\midrule
Train & 4.18 s & 3.23 s & 92.84 s & 96.24 s \\
Test & 4.06 s & 3.25 s & 82.68 s & 87.91 s \\
\bottomrule
\end{tabular}}
\end{table}

\begin{table}[t]
\centering
\small
\caption{Emotion label distribution of StreamMER.}
\label{tab:label-distribution}
\begin{tabular}{lrr}
\toprule
Emotion & Train & Test \\
\midrule
Joy & 2,405 & 249 \\
Anger & 1,042 & 152 \\
Neutral & 1,315 & 152 \\
Anxiety & 1,339 & 83 \\
Surprise & 642 & 97 \\
Sadness & 733 & 33 \\
Embarrassment & 402 & 48 \\
Disgust & 233 & 1 \\
\bottomrule
\end{tabular}
\end{table}

\paragraph{Streaming evaluation split.}
The test split contains 815 utterances with available prefix-video assets. These samples are used for streaming-prefix evaluation, where Online Prefix Belief Formation only observes partial audio prefixes and Contextual Belief Reinterpretation is requested selectively when the belief trajectory is judged unreliable.

\begin{table}[t]
\centering
\small
\caption{Statistics of StreamMER.}
\label{tab:dataset-statistics-compact}
\begin{tabular}{lccc}
\toprule
Split & Utterances & Clips & Episodes \\
\midrule
Train & 8,111 & 365 & 48 \\
Test & 815 & 40 & 26 \\
\bottomrule
\end{tabular}
\end{table}

\section{Additional Analysis of Belief Trajectories}
\label{sec:belief-trajectory-analysis}

\begin{figure}[t]
\centering
\includegraphics[width=\columnwidth]{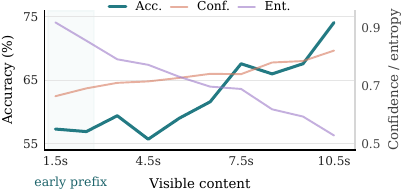}
\caption{Prefix-level belief dynamics under increasing observation length. Confidence increases and entropy decreases as more content is observed, while accuracy is not strictly monotonic.}
\label{fig:prefix_acc}
\end{figure}

Figure~\ref{fig:prefix_acc} analyzes prefix-level belief dynamics under increasing observation length.
As more content is observed, the model's average confidence generally increases and entropy decreases, indicating that the predicted distribution becomes more concentrated.
However, accuracy is not strictly monotonic.
At some intermediate observation lengths, confidence continues to increase while accuracy drops.
This suggests that high confidence at a single prefix does not necessarily imply a reliable prediction; the model also needs to determine whether the confidence is formed through stable evidence accumulation or caused by short-term fluctuations.

Table~\ref{tab:switch} further shows that belief trajectories contain dynamic reliability information beyond single-prefix predictions.
When the predicted label does not switch, Online Prefix Belief Formation reaches 72.6\% accuracy.
In contrast, the accuracy drops to 38.3\% with one switch and 24.1\% with three switches.
Meanwhile, Contextual Belief Reinterpretation provides larger correction benefits for unstable trajectories.
For samples with one switch, reinterpretation produces 74 helpful corrections and only 19 harmful changes.
This indicates that trajectory instability reflects not only the error risk of the current belief, but also the potential benefit of introducing additional contextual evidence.

\begin{table}[t]
\centering
\small
\caption{Prediction-switch count and trajectory reliability on the matched validation subset. Accuracy values are percentages; Gain+ and Gain- denote helpful and harmful contextual reinterpretations.}
\label{tab:switch}
\resizebox{\columnwidth}{!}{
\begin{tabular}{rrrrr r}
\toprule
Switches & Samples & Belief Acc. & Reinterp. Acc. & Gain+ & Gain- \\
\midrule
0 & 467 & 72.6 & 75.6 & 40 & 26 \\
1 & 222 & 38.3 & 63.1 & 74 & 19 \\
2 & 72 & 44.4 & 68.1 & 22 & 5 \\
3 & 29 & 24.1 & 48.3 & 9 & 2 \\
\bottomrule
\end{tabular}}
\end{table}

Overall, these results show that belief trajectories are not merely intermediate prediction records, but useful dynamic reliability signals.
Compared with relying only on single-prefix confidence, trajectory information reveals whether an emotion belief is formed stably, whether competing labels remain unresolved, and whether Contextual Belief Reinterpretation is needed.

\section{Effect of Context Scope in Contextual Belief Reinterpretation}
\label{sec:context-scope}

We evaluate how much contextual evidence should be used in Contextual Belief Reinterpretation on StreamMER. This experiment isolates the third module and directly compares three context scopes: using only the current target video, using the previous video together with the current target video, and using the previous, current, and next videos. The goal is to examine whether enlarging the contextual window consistently improves emotion reinterpretation. The Previous + Current setting is the streaming-compatible default used by TRACE. The Previous + Current + Next setting is included only as an offline diagnostic comparison, since future utterances are not available under the online streaming protocol.

\begin{table}[t]
\centering
\small
\caption{Recognition performance of Contextual Belief Reinterpretation with different context scopes on StreamMER.}
\label{tab:context-scope-performance}
\resizebox{\columnwidth}{!}{
\begin{tabular}{lccc}
\toprule
Context Scope & Acc. & W-F1 & UAR \\
\midrule
Current & 62.53 & 62.08 & 50.26 \\
Previous + Current & \textbf{70.18} & \textbf{70.00} & \textbf{58.29} \\
Previous + Current + Next & 68.83 & 68.56 & 55.83 \\
\bottomrule
\end{tabular}}
\end{table}

Table~\ref{tab:context-scope-performance} shows that contextual scope has a substantial effect on reinterpretation quality. Using only the current target video obtains 62.53 Accuracy, indicating that the current clip alone is often insufficient for resolving ambiguous emotional expressions. When the previous video is added, performance improves markedly to 70.18 Accuracy, 70.00 W-F1, and 58.29 UAR. This suggests that preceding dialogue context provides important cues about the speaker's emotional state, especially when the current utterance is short, implicit, or emotionally dependent on earlier interaction.

Adding the next video does not further improve performance. The Previous + Current + Next setting reaches 68.83 Accuracy, which is lower than the Previous + Current setting. This result suggests that more context is not always beneficial. Future context can introduce distracting evidence, speaker shifts, or emotional transitions that are not directly relevant to the target utterance. It is also incompatible with the online streaming protocol because future information is not immediately available. Therefore, we use Previous + Current as the default context scope for Contextual Belief Reinterpretation, and report Previous + Current + Next only as an offline diagnostic comparison.

\begin{table}[t]
\centering
\small
\caption{Inference efficiency of Contextual Belief Reinterpretation with different context scopes. Avg. latency denotes the average wall-clock inference time per sample. Video sec/s denotes how many seconds of video can be processed per second of computation. RTF is the real-time factor, computed as inference time divided by video duration; values below 1 indicate real-time processing.}
\label{tab:context-scope-speed}
\resizebox{\columnwidth}{!}{
\begin{tabular}{lccc}
\toprule
Context Scope & Avg. Latency (s) & Video sec/s & RTF \\
\midrule
Current & 0.3636 & 11.2281 & 0.0891 \\
Previous + Current & 0.8959 & 4.8809 & 0.2049 \\
Previous + Current + Next & 0.8999 & 4.8592 & 0.2058 \\
\bottomrule
\end{tabular}}
\end{table}

Table~\ref{tab:context-scope-speed} reports the inference efficiency of the three context scopes. The current-only setting is the fastest because it processes only the target video. Adding the previous video increases the average latency from 0.3636s to 0.8959s, reflecting the additional cost of using contextual evidence. However, the throughput remains 4.8809 video seconds per second, and the RTF is 0.2049, which is still well below 1. This means that the Previous + Current setting can process video faster than real time and is compatible with streaming deployment.

The Previous + Current + Next setting has nearly the same computational cost as Previous + Current, with 0.8999s average latency and 0.2058 RTF, but it performs worse in recognition. This offline diagnostic result shows that its limitation is not mainly computational cost, but the quality and relevance of the added future context. Combining the performance and efficiency results, Previous + Current gives the best overall trade-off: it substantially improves recognition over Current while preserving real-time inference and avoiding the instability introduced by future context.

\section{Additional Robustness and Generalization Analyses}
\label{app:robustness}

\subsection{Fast-Path Modality Choice}

The audio-first online pathway is motivated by latency--utility trade-offs
rather than by an assumption that audio is universally more informative
than vision for emotion understanding. We therefore compare audio-only,
visual-only, and audio--visual inputs for the fast pathway while keeping
the remaining TRACE pipeline unchanged.

\begin{table*}[t]
\centering
\small
\setlength{\tabcolsep}{4pt}
\begin{tabular}{lccccccccc}
\toprule
& \multicolumn{4}{c}{Online Path}
& \multicolumn{5}{c}{TRACE} \\
\cmidrule(lr){2-5}
\cmidrule(lr){6-10}
Fast-Path Modality
& Acc. & W-F1 & UAR & RTF
& Acc. & W-F1 & UAR & RR & RTF \\
\midrule
Audio
& 59.87 & 54.56 & 52.58 & 0.055
& 69.47 & 68.10 & 63.80 & 55.58 & 0.091 \\

Visual
& 42.45 & 36.26 & 39.84 & 0.081
& 66.26 & 64.95 & 59.39 & 64.17 & 0.113 \\

Audio+Visual
& 60.74 & 54.31 & 52.86 & 0.083
& \textbf{69.57} & 67.96 & \textbf{64.02} & 56.07 & 0.119 \\
\bottomrule
\end{tabular}
\caption{
Effect of fast-path modality choice on StreamMER.
RR denotes contextual reinterpretation rate.
Audio provides the strongest latency--utility trade-off for early belief
formation, while adding visual input to every prefix yields only marginal
accuracy gains at higher inference cost.
}
\label{tab:modality_ablation}
\end{table*}

Adding visual evidence to every online prefix increases Online Path accuracy
only from 59.87\% to 60.74\%, while increasing RTF from 0.055 to 0.083.
For the complete TRACE system, the corresponding accuracy improvement is
only 0.10 points (69.47\% to 69.57\%), whereas RTF increases from 0.091
to 0.119. These results support using audio as the low-latency primary
signal while reserving richer multimodal evidence for samples whose belief
trajectories are judged unreliable. Importantly, this result should not be
interpreted as claiming that audio is intrinsically superior to vision for
emotion recognition.

\subsection{Robustness to Acoustic Noise}

We further evaluate whether trajectory-aware routing can adapt when the
reliability of the primary acoustic signal degrades. We apply the same
10-dB acoustic-noise setting used in our robustness evaluation and compare
the online pathway with the complete TRACE system.

\begin{table}[t]
\centering
\small
\begin{tabular}{lccc}
\toprule
Condition & Online Acc. & TRACE Acc. & TRACE RR \\
\midrule
Clean & 59.87 & 69.47 & 55.58 \\
10-dB Noise & 53.62 & 64.66 & 68.22 \\
\bottomrule
\end{tabular}
\caption{
Robustness under acoustic noise on StreamMER.
As the acoustic evidence becomes less reliable, TRACE increases the
contextual reinterpretation rate automatically.
}
\label{tab:noise_robustness}
\end{table}

Acoustic corruption reduces the online-path accuracy from 59.87\% to
53.62\%. TRACE partially recovers this degradation, reaching 64.66\%
accuracy. At the same time, the reinterpretation rate automatically rises
from 55.58\% to 68.22\%, indicating that degraded acoustic evidence is
reflected in less stable belief trajectories and consequently triggers
richer contextual reasoning more frequently.

\section{Ablation of Rationale and Memory Summary}
\label{sec:rationale-summary-ablation}

We further analyze the role of two output components in Contextual Belief Reinterpretation: the classification rationale and the contextual memory summary.
In the default TRACE reinterpretation model, the model predicts the final emotion label, generates a brief rationale, and updates a compact memory summary for subsequent utterances.
The rationale encourages the model to ground its prediction in observable audio-visual and dialogue evidence, while the memory summary preserves preceding conversational state for later reinterpretation.

\begin{table}[t]
\centering
\small
\caption{Ablation of rationale generation and contextual memory summary in Contextual Belief Reinterpretation on StreamMER. Values are percentages.}
\label{tab:rationale-summary-ablation}
\resizebox{\columnwidth}{!}{
\begin{tabular}{lccc}
\toprule
Setting & Acc. & W-F1 & UAR \\
\midrule
w/o Memory Summary & 66.01 & 65.69 & 61.81 \\
w/o Rationale & 63.31 & 61.84 & 55.31 \\
TRACE & 70.18 & 70.00 & 58.29 \\
\bottomrule
\end{tabular}}
\end{table}

Table~\ref{tab:rationale-summary-ablation} shows that both rationale generation and memory summary contribute to contextual reinterpretation.
Removing the memory summary reduces Accuracy from 70.18\% to 66.01\% and W-F1 from 70.00\% to 65.69\%, indicating that preceding conversational state helps resolve emotion ambiguity in the current utterance.
Interestingly, this variant still obtains a relatively high UAR of 61.81\%, suggesting that it may preserve some minority-class sensitivity but loses overall recognition reliability.

Removing rationale generation causes a larger degradation, reducing Accuracy to 63.31\%, W-F1 to 61.84\%, and UAR to 55.31\%.
This result suggests that requiring the model to produce an explicit rationale helps it ground the final emotion label in audio-visual cues, dialogue semantics, and contextual evidence.
Overall, the ablation indicates that the strongest reinterpretation performance comes from jointly predicting the emotion label, explaining the decision, and maintaining a compact contextual memory.

\section{Implementation Details}
\label{sec:implementation-details}

\paragraph{Baseline implementation.}
For a fair comparison, all trainable baselines are fine-tuned on the corresponding training split of each dataset and evaluated on the same held-out test split and label space. On StreamMER, we provide each baseline with the same available utterance-level evidence and contextual scope used by TRACE whenever the model architecture supports the required modalities. For models that do not support a specific modality, we use their strongest compatible input setting following the original implementation. Closed-source models are evaluated with the same prompt format, label definitions, and available evidence. No baseline is tuned on the test split.

\paragraph{TRACE implementation.}
The online prefix belief formation module uses Qwen2.5-Omni-3B as the backbone, while the contextual belief reinterpretation module uses Qwen2.5-Omni-7B. Both modules are parameter-efficiently fine-tuned with LoRA for 10 epochs. The LoRA rank is set to 16, with $\alpha=32$ and a dropout rate of 0.05. LoRA is applied to the attention and feed-forward projections, including \texttt{q\_proj}, \texttt{k\_proj}, \texttt{v\_proj}, \texttt{o\_proj}, \texttt{gate\_proj}, \texttt{up\_proj}, and \texttt{down\_proj}. Training is conducted with mixed precision and distributed over 8 GPUs using DeepSpeed. We use AdamW with a learning rate of $8\times10^{-6}$, weight decay of 0.01, warmup ratio of 0.05, and a maximum gradient norm of 1.0.

\paragraph{Online prefix belief formation.}
Audio prefixes are constructed starting from 1.5 seconds and are extended with a stride of 1.0 second. The complete utterance is also retained as a full-prefix training sample. Audio is resampled to 16 kHz. During training, all prefix predictions are supervised by the emotion label of the current utterance, encouraging the model to form useful early emotion beliefs under incomplete observations. The per-device training batch size is set to 4, with gradient accumulation steps set to 1.

\paragraph{Contextual belief reinterpretation.}
The contextual belief reinterpretation module is triggered to make a revised prediction by integrating the multimodal information of the current utterance, preceding contextual information, and the belief trajectory produced during the online stage. Visual frames are sampled at 2 fps, with the longer image side resized to 448. The streaming-compatible input includes the current video segment and its preceding contextual video, which are sampled in the same way. Following contextual videos are used only in the offline diagnostic context-scope analysis, not in the default TRACE streaming protocol. The module outputs the final emotion label, an explanation, and an updated contextual summary. The per-device training batch size is 2, the gradient accumulation steps are 2, and the per-device evaluation batch size is 8.

\paragraph{Trajectory-calibrated reliability estimation.}
The trajectory-calibrated reliability estimation module trains a trigger based on the belief trajectory generated during the online prefix stage. The trigger features include final confidence, entropy, inter-class margin, changes in confidence, entropy, and margin between adjacent prefixes, JS divergence, top-1 persistence, entropy AUC, the number of class switches, the number of local switches, observation ratio, final probabilities for different emotion classes, and log-likelihood features. We adopt a two-head risk-gain model to jointly predict the risk of an incorrect online prefix prediction and the expected gain of invoking the reinterpretation module. The model uses a hidden dimension of 128, dropout of 0.3, and a batch size of 128. It is optimized with AdamW using a learning rate of $1\times10^{-3}$ and weight decay of $1\times10^{-3}$. A cosine learning-rate schedule is used. The model is trained for at most 300 epochs, with early stopping if the validation loss does not improve for 30 consecutive epochs.

\paragraph{Leakage control and checkpoint selection.}
To prevent leakage, trajectory features and risk/gain labels for reliability training are generated with $K$-fold out-of-fold predictions on the training split. No test example is used for fitting the reliability model, selecting thresholds, calibrating the policy, or choosing hyperparameters. All experiments use the same predefined train/test split for StreamMER and the official or predefined splits for MELD and MER2024. The online prefix belief formation and contextual belief reinterpretation modules are trained with LoRA for a fixed maximum number of epochs, and the best checkpoint is selected according to validation performance. All reported results are computed on the held-out test split using the selected checkpoint.

\paragraph{Inference and evaluation.}
For generative emotion prediction, we use deterministic decoding with temperature set to 0, so that model outputs are reproducible under the same checkpoint and input. For classification from label probabilities, the predicted label is selected by greedy decoding or by taking the emotion label with the highest normalized probability within the predefined label space. Unless otherwise specified, we report the result of the selected checkpoint under deterministic inference rather than averaging stochastic decoding samples. For latency and throughput evaluation, all methods are measured under the same hardware and inference configuration, and we report per-sample latency and real-time factor using the same evaluation split.

\newtcblisting{promptbox}{
    colback=gray!5,
    colframe=gray!55,
    arc=3mm,
    boxrule=0.6pt,
    left=2mm,
    right=2mm,
    top=2mm,
    bottom=2mm,
    width=\columnwidth,
    breakable,
    listing only,
    listing options={
        basicstyle=\ttfamily\scriptsize,
        breaklines=true,
        columns=fullflexible,
        keepspaces=true
    }
}

\section{Feature Details}
\label{sec:feature-details}

This appendix describes the trajectory features used as the input of Trajectory-Calibrated Reliability Estimation. For each utterance $V_k$, Online Prefix Belief Formation produces a sequence of belief distributions
\begin{equation}
\mathcal{B}_{k,1:t}
=
\{p_{k,1},p_{k,2},\ldots,p_{k,t}\},
\end{equation}
where $p_{k,i}\in\mathbb{R}^{|\mathcal{Y}|}$ is the emotion probability distribution at prefix step $i$. The trigger feature vector $z_{k,t}=\phi(\mathcal{B}_{k,1:t})$ is computed from the current belief, its temporal changes, and the stability of the predicted label sequence.

\paragraph{Current belief features.}
We first extract features from the current prefix distribution $p_{k,t}$. The confidence is defined as the maximum class probability:
\begin{equation}
c_t=\max_{y\in\mathcal{Y}}p_{k,t}(y).
\end{equation}
The predicted label is
\begin{equation}
\hat{y}_{k,t}^{\mathrm{on}}
=
\arg\max_{y\in\mathcal{Y}}p_{k,t}(y).
\end{equation}
We also compute the entropy of the current distribution:
\begin{equation}
H_t
=
-\sum_{y\in\mathcal{Y}}
p_{k,t}(y)\log p_{k,t}(y),
\end{equation}
and the inter-class margin between the largest and second-largest probabilities:
\begin{equation}
m_t=p_{k,t}^{(1)}-p_{k,t}^{(2)},
\end{equation}
where $p_{k,t}^{(1)}$ and $p_{k,t}^{(2)}$ denote the top-1 and top-2 class probabilities. A high confidence and a large margin usually indicate a sharp belief, while high entropy indicates uncertainty.

\paragraph{Temporal change features.}
To capture whether the belief is becoming more reliable over time, we compute changes between adjacent prefixes:
\begin{equation}
\begin{aligned}
\Delta c_i &= c_i-c_{i-1},\\
\Delta H_i &= H_i-H_{i-1},\\
\Delta m_i &= m_i-m_{i-1}.
\end{aligned}
\end{equation}
From these sequences, we use summary statistics such as the latest change, mean change, and maximum absolute change. These features describe whether the belief is gradually stabilizing or still fluctuating.

\paragraph{Distributional stability features.}
We measure the stability of consecutive belief distributions using Jensen-Shannon divergence:
\begin{equation}
\begin{aligned}
\mathrm{JS}(p_{k,i},p_{k,i-1})
=&\frac{1}{2}\mathrm{KL}(p_{k,i}\|u_i) \\
&+\frac{1}{2}\mathrm{KL}(p_{k,i-1}\|u_i),
\end{aligned}
\end{equation}
where
\begin{equation}
u_i=\frac{1}{2}(p_{k,i}+p_{k,i-1}).
\end{equation}
We use summary statistics of adjacent JS divergence values over the observed trajectory, including the mean and maximum divergence. Larger divergence indicates that the belief distribution changes substantially after new evidence arrives.

\paragraph{Label-switching features.}
We also compute features from the predicted label sequence
\begin{equation}
\hat{Y}_{k,1:t}^{\mathrm{on}}
=
\{\hat{y}_{k,1}^{\mathrm{on}},
\ldots,
\hat{y}_{k,t}^{\mathrm{on}}\}.
\end{equation}
The switch count is defined as
\begin{equation}
S_t
=
\sum_{i=2}^{t}
\mathbb{I}
\left(
\hat{y}_{k,i}^{\mathrm{on}}
\neq
\hat{y}_{k,i-1}^{\mathrm{on}}
\right).
\end{equation}
We further compute top-1 persistence, namely the fraction of observed prefixes whose predicted label matches the current top-1 label:
\begin{equation}
P_t
=
\frac{1}{t}
\sum_{i=1}^{t}
\mathbb{I}
\left(
\hat{y}_{k,i}^{\mathrm{on}}
=
\hat{y}_{k,t}^{\mathrm{on}}
\right).
\end{equation}
These features capture whether the current prediction is supported by a stable trajectory or emerges after repeated label changes.

\paragraph{Trajectory accumulation features.}
To summarize uncertainty over the whole observed trajectory, we compute the area under the entropy curve:
\begin{equation}
\mathrm{AUC}_H(t)
=
\frac{1}{t}\sum_{i=1}^{t}H_i.
\end{equation}
A high entropy AUC means that the model remains uncertain for most of the observed prefixes, even if the final prefix becomes confident.

\paragraph{Observation progress features.}
Finally, we include the observation ratio
\begin{equation}
r_t=\frac{t}{T_k},
\end{equation}
where $T_k$ is the total number of prefix steps for utterance $V_k$. This feature informs the trigger whether the current belief is formed from an early partial observation or from a nearly complete utterance.

\paragraph{Class-probability and likelihood features.}
In addition to the aggregated trajectory statistics above, we include the current class probability vector $p_{k,t}$ itself. We also include log-probability features for the top predicted classes:
\begin{equation}
\log p_{k,t}(y),
\end{equation}
with numerical clipping for stability. These features preserve class-specific information that may be useful for estimating whether contextual reinterpretation is likely to help.

Overall, the trigger input combines current confidence, entropy, margin, temporal changes, JS divergence, label-switching statistics, top-1 persistence, entropy AUC, observation ratio, final class probabilities, and log-likelihood features. These features allow the reliability model to distinguish a confident and stable belief from a confident but unstable one, which is crucial for deciding whether to commit immediately or request Contextual Belief Reinterpretation.

\section{Prompt Templates}
\label{app:prompt-templates}

This appendix provides the prompt templates used in StreamMER construction and TRACE inference.
The StreamMER annotation prompts are used to generate initial utterance-level emotion annotations before manual verification.
The Online Prefix Belief Formation prompt is used by the fast streaming belief tracker to estimate emotion beliefs from observed audio prefixes.
The Contextual Belief Reinterpretation prompt is used when the trajectory-calibrated policy requests additional multimodal reasoning for uncertain samples.
All prompts are shown verbatim.

\subsection{Online Prefix Belief Formation Prompt}

\begin{promptbox}
You are the fast belief tracker in streaming emotion understanding.
Infer the speaker's current emotion from the observed prefix only.
The evidence is partial and arrives incrementally.
Prioritize vocal cues such as prosody, speaking rate, pauses, tremble, breathiness, laughter, crying, and intensity.
Do not assume evidence from the unobserved future part of the clip.
Observed prefix duration: {prefix_sec:.2f}s out of total clip duration {clip_duration_sec:.2f}s.
Observed fraction of the clip: {observed_ratio:.2%}.
Reply with exactly one emotion label from: neutral, anger, anxiety, sadness, joy, surprise, embarrassment.
The speaker's emotion in the observed prefix is:
\end{promptbox}

\subsection{Contextual Belief Reinterpretation Prompt}

\begin{promptbox}
Task: predict the target speaker's emotion in the current clip.

You will receive up to three multimodal inputs in this order:
1. optional previous-context audio
2. the current target audio clip
3. the current target video clip

The current target audio and current target video describe the same clip and are the primary evidence.

Prioritize the current target audio when judging subtle emotions, especially prosody, speaking rate, pauses, intensity, breathiness, pitch movement, trembling, laughter, crying, sighs, and voice quality changes.

Use the current target video to verify or refine the emotion from facial expression, gaze, posture, gesture, and visible interaction context.

The previous-context audio is auxiliary context only. Use it only to resolve ambiguity in the current clip, never as the main basis for prediction. Never predict the emotion of the previous clip.

Valid emotion labels:
neutral, anger, anxiety, sadness, joy, surprise, embarrassment.

Return the final emotion label for the current clip.

Target speaker in the current clip: {speaker}


History summary before the current clip:
"{history_summary}"

Use the history summary as background state, not as a replacement for current audio-visual evidence.

Reference belief trajectory before reinterpretation:
- Fast prediction: {fast_pred}
- Fast probabilities: {fast_probs}
- Fast entropy: {fast_entropy}
- Fast margin: {fast_margin}
- Recent fast labels: {traj_last_k}

The reference belief trajectory is provided only to indicate why the online belief may be uncertain or unstable. It is not a final answer. You may accept or reject it based on the current audio-visual evidence and the previous-context audio.

When audio and video disagree, trust the current target audio more for affective state unless the audio is clearly off-screen, corrupted, silent, or belongs to another speaker.

Focus on the target speaker's own voice rather than background music, sound effects, or other speakers.

Output strict JSON with keys:
- final_emotion: one label from the valid set
- reason: a brief explanation grounded in current audio, current video, dialogue, or previous context
- new_summary: a concise summary of the current clip for updating dialogue memory
\end{promptbox}

\subsection{StreamMER Annotation Prompts}

\paragraph{User prompt.}

\begin{promptbox}
USER_PROMPT = """
Analyze the uploaded video together with the [Reference Script].

The video provides emotional evidence, including facial expression, voice, prosody, body language, gaze, and local interaction context.
The [Reference Script] provides the exact timestamps, speakers, and dialogue text.

Use the script as the segmentation anchor and the video as the emotional evidence.

The video may contain multiple scenes. Analyze all dialogue segments in the script.

For EACH dialogue segment, output one JSON object with the following fields:

{
  "timestamp": {
    "start": "Copy EXACTLY from [Reference Script], including brackets",
    "end": "Copy EXACTLY from [Reference Script], including brackets"
  },
  "speaker": "Copy EXACTLY from [Reference Script]",
  "dialogue": "Copy dialogue text from [Reference Script], or concatenate merged lines if subtitle splitting occurs",
  "final_emotion_label": "Choose exactly one from: Joy, Sadness, Anger, Surprise, Anxiety, Disgust, Embarrassment, Neutral",
  "local_summary": "A short emotion-focused summary of what is happening in this utterance and its immediate local context",
  "reason": "A concise explanation for the emotion label, grounded in audio, visual, dialogue, and local context evidence"
}

Additional requirements:
- The reason must mention concrete evidence from the video when available, such as tone, pitch, speaking rate, hesitation, facial expression, gaze, posture, gesture, or body movement.
- The local_summary should focus on emotion-relevant context only. Do not summarize unnecessary plot logistics.
- If the speaker is off-screen or visual evidence is unavailable, rely on audio and dialogue, and state that visual evidence is limited.
- If audio and visual cues conflict, choose the final label based on the stronger evidence and explain briefly in reason.
- If emotional evidence is weak or ambiguous, choose Neutral conservatively.
- Do not add fields beyond the specified schema.
- Output the full JSON list in minified format only.

[Reference Script]:
{script_content}

Output the JSON list now.
"""
\end{promptbox}

\paragraph{System prompt.}

\begin{promptbox}
SYSTEM_PROMPT = """
You are a multimodal emotion annotation assistant.

Use BOTH the uploaded video and the [Reference Script] to create utterance-level emotion annotations.

CRITICAL RULES:
- Use the video as the primary source for emotional evidence, including facial expression, tone, prosody, body language, gaze, and interaction context.
- Use the [Reference Script] as the ONLY source for timestamps, speaker names, and dialogue text.
- The fields timestamp.start, timestamp.end, speaker, and dialogue MUST copy the script exactly.
- Timestamps must keep the original bracket format, e.g. [00:02:42:440].
- Never rewrite, normalize, or invent timestamps.
- Do not use external knowledge about the TV show.

SEGMENTATION RULE:
Annotate each dialogue segment according to the script.
Sometimes subtitles split one spoken sentence into multiple lines.
You MAY merge consecutive script lines into one annotation object only if:
- the speaker is the same,
- the lines clearly form one continuous sentence or emotional unit,
- no other speaker appears between them,
- the merged segment is not too long, ideally within 10 seconds,
- no more than 3 consecutive lines are merged.

When merging:
- timestamp.start = start timestamp of the first merged line,
- timestamp.end = end timestamp of the last merged line,
- dialogue = concatenation of the merged dialogue text.

Do NOT merge across different speakers or conversational turns.
If unsure, keep lines separate.

ANNOTATION GOAL:
For each utterance-level segment, identify the speaker's dominant observable core emotion and provide concise evidence-based annotation.

LABEL RULE:
final_emotion_label must represent the CORE emotion, not merely speaking style or social expression.
Valid labels are:
Joy, Sadness, Anger, Surprise, Anxiety, Disgust, Embarrassment, Neutral

If sarcasm, teasing, masking, or politeness exists, infer the underlying core emotion when the video evidence supports it.
Examples:
- playful teasing -> Joy
- sarcastic anger -> Anger
- nervous politeness -> Anxiety
- weak or unclear emotion -> Neutral

OUTPUT FORMAT:
Output ONLY a raw, minified, single-line JSON list.
Do not output markdown, explanations, comments, or extra text.
"""
\end{promptbox}

\section{Complete Prefix Spectrogram Visualization}
\label{sec:prefix-visualization}

\begin{figure*}[t]
\centering
\includegraphics[width=\textwidth]{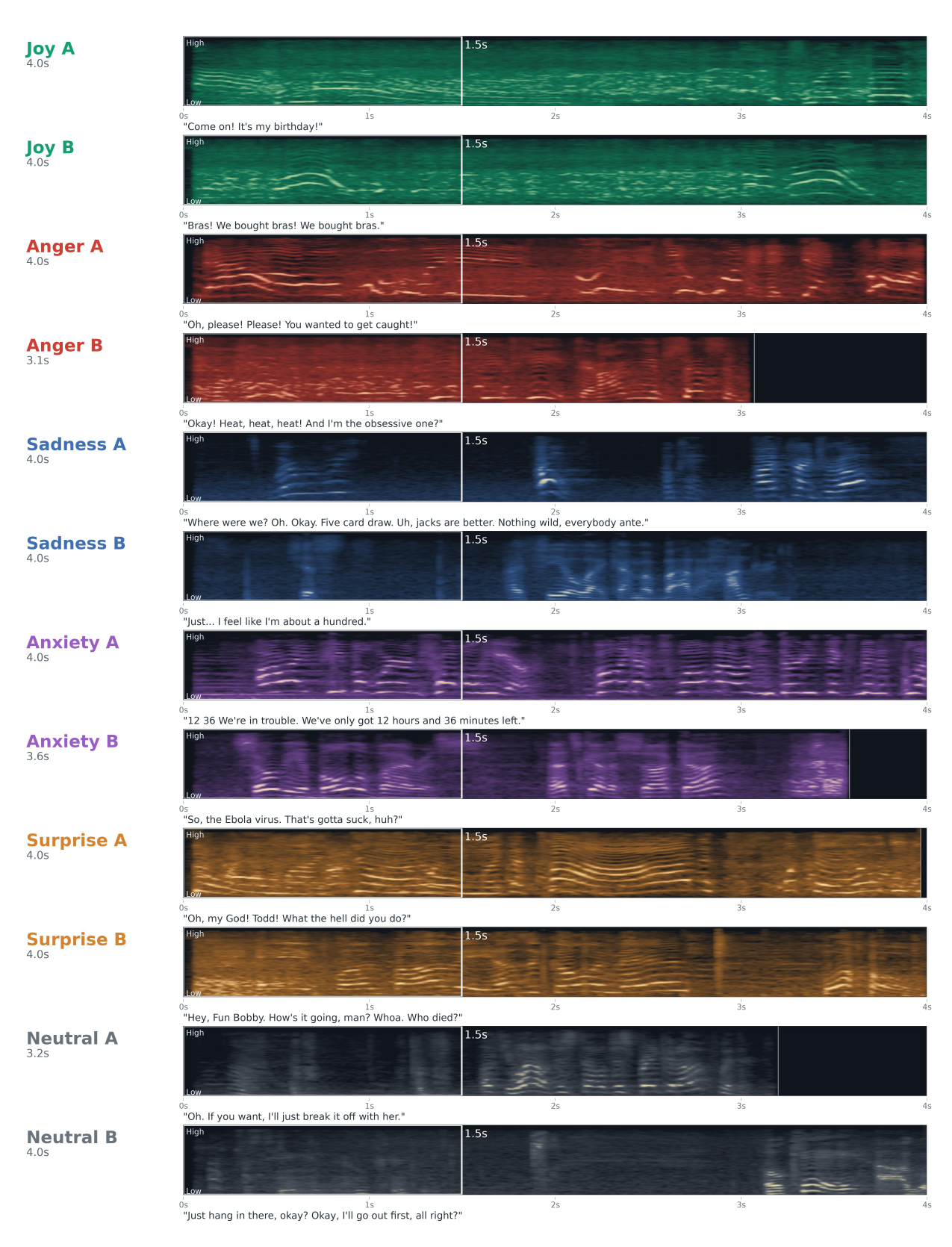}
\caption{Complete spectrogram visualization of early audio prefixes across emotion categories.}
\label{fig:prefix-visualization-full}
\end{figure*}

Figure~\ref{fig:prefix-visualization-full} provides a more complete visualization of acoustic evidence in early audio prefixes across emotion categories.
These spectrograms complement Figure~\ref{fig:prefix_show} in the main text by showing that short prefix observations already contain category-dependent acoustic patterns, including differences in energy distribution, spectral structure, and temporal rhythm.
The visualization supports the use of streaming audio prefixes as the primary low-latency signal for Online Prefix Belief Formation.

\end{document}